\documentclass[12pt]{article}
\usepackage{amsfonts,amsthm,amsmath,amssymb,upgreek,bm}
\usepackage{hyperref}
\usepackage[paper=letterpaper,margin=.85in]{geometry}
\usepackage{graphicx}
\usepackage{units}
\usepackage{float}
\usepackage[export]{adjustbox}
\usepackage{xcolor}
\usepackage[shortlabels]{enumitem}
\usepackage[mathscr]{euscript}
\usepackage[normalem]{ulem}
\usepackage{bm}
\usepackage{tikz}
\usepackage{tikzsymbols}
\usepackage{pgfplots}
\usepackage{ytableau}
\usepackage{slashed}
\usepackage{mathtools}
\usepackage{bbold}
\usetikzlibrary{decorations.pathmorphing}
\usetikzlibrary{arrows.meta}
\usetikzlibrary{calc}

\newcommand{\be}{\begin{equation}}
\newcommand{\ee}{\end{equation}}
\newcommand{\bea}{\begin{eqnarray}}
\newcommand{\eea}{\end{eqnarray}}
\renewcommand{\tilde}{\widetilde}
\renewcommand{\i}{\mathrm{i}}
\renewcommand{\d}{\mathrm{d}}
\newcommand{\x}{\mathsf{x}}

\newcommand{\K}{\mathbb{K}}
\newcommand{\Z}{\mathbb{Z}}
\newcommand{\one}{\mathbb{1}}

\numberwithin{equation}{section}

\def\tr{\text{Tr}}

\begin{document}
\thispagestyle{empty}

\vspace*{2.5cm}
\begin{center}

{\bf {\LARGE Cusps in 3d gravity}}

\begin{center}

\vspace{1cm}

{Douglas Stanford${}^1$ and Cynthia Yan${}^{1,2}$}\\
 \bigskip \rm

\bigskip 

{${}^1$  Leinweber Institute for Theoretical Physics, Stanford University, Stanford, CA 94305, USA}
\,\vspace{0.5cm}\\
{${}^2$ Department of Physics, Harvard University, Cambridge, MA 02138, USA}

\rm
  \end{center}

\vspace{2.5cm}
{\bf Abstract}
\end{center}
\begin{quotation}
\noindent

Three dimensional hyperbolic manifolds have accumulation points in the spectrum of their volumes, leading to a divergence in the sum over topologies. The limit points are cusped hyperbolic manifolds, and we propose to renormalize the sum by including the cusped manifold as a counterterm. This gives a reinterpretation of the zeta-function regularization procedure used by Maloney and Witten in the sum over SL$(2,\mathbb{Z})$ black holes. For pure $\mathcal{N} = 1$ supergravity, cusps with even spin structure can be used in a similar way. Cusps with odd spin structure are not needed to cancel any divergence, but they find an application by making the index nonzero.

\end{quotation}

\setcounter{page}{0}
\setcounter{tocdepth}{2}
\setcounter{footnote}{0}

\newpage

\parskip 0.1in
 
\setcounter{page}{2}
\tableofcontents

\newpage

\section{Introduction}

In the path integral approach to quantum gravity, the goal is to sum over the topology and geometry of $d$-dimensional manifolds $M$ with fixed boundary conditions $\partial M$:
\be\label{qg:label}
Z(\partial M) = \sum_{\text{topology}(M)}\int_{\text{geometry}(M)} e^{-I(M)}.
\ee
To what extent does this make sense, especially the sum over topologies? 

In $d = 2$ dimensions, the sum makes sense as asymptotic series, because topologies have an ordering by their integer Euler characteristic $\chi$. If we include $\chi$ in the action with a large negative coefficient, complicated topologies will be suppressed. 

We will focus on the case $d = 3$, and we will take $I$ to be the Einstein-Hilbert action for gravity with a negative cosmological constant
\be{}
I = -\frac{1}{16\pi G}\left[\int_M \sqrt{g}\left(R + 2\right) +2\int_{\partial M}\sqrt{h}(K-1)\right].
\ee
For small $G$, the integral over geometries with a given topology will be dominated by a saddle point (if one exists) satisfying Einstein's equations $R_{\mu\nu} - \frac{1}{2}g_{\mu\nu}R = g_{\mu\nu}$. Such a saddle point is a hyperbolic manifold, for which the action reduces to\footnote{For a closed manifold $\text{vol}$ is literally the hyperbolic volume. For manifold with an asymptotic boundary, this is generalized to the renormalized volume, see e.g.~\cite{Schlenker:2022dyo}.}
\be\label{EHaction:label}
I = \frac{\text{vol}(M)}{4\pi G}.
\ee
So saddle points are weighted by $e^{-\#\text{vol}(M)}$, and big complicated manifolds are suppressed just like in two dimensions. Indeed, (minus) the hyperbolic volume can be considered a type of generalization of the Euler characteristic to three dimensions.

However, there are some problems. For one thing, there are infinitely many topologies with no hyperbolic metric. These ``off shell topologies'' should also contribute to (\ref{qg:label}), but tools to compute them are still under development.\footnote{See \cite{Maxfield:2020ale,Cotler:2020ugk,Yan:2023rjh,Jafferis:2024jkb,Post:2024itb,Yan:2025usw,Boruch:2025ilr,deBoer:2025rct,Hartman:2025ula} for some ideas and puzzles.} For a second thing, in three dimensions (and only three \cite{gromov1981hyperbolic}) the spectrum of hyperbolic volumes has accumulation points. In particular, there are infinitely many topologies with hyperbolic volume less than a certain finite value.\footnote{The first accumulation point for closed orientable manifiolds is at $\text{vol}(\text{figure 8 complement}) \approx 2.02988$.} So the suppression by the volume is not enough to make (\ref{qg:label}) an asymptotic series.\footnote{It might be that the first problem resolves the second: maybe the contribution of off-shell manifolds cancels the divergence from the sum over accumulation points of hyperbolic manifolds. That would be wonderful, but in this paper we will propose a way to make sense of (\ref{qg:label}) if this miracle does not occur.}

This problem was encountered in \cite{Maldacena:1998bw,Dijkgraaf:2000fq,Maloney:2007ud}, and in \cite{Maloney:2007ud} Maloney and Witten defined the divergent sum near an accumulation point using a version of zeta function regularization. In this paper we will reinterpret the Maloney/Witten prescription slightly. In particular, we introduce new contributions to (\ref{qg:label}) from manifolds with cusps, and we adjust the fugacity for cusps to cancel the divergence from the accumulation points. Our analysis is conditional on an assumption that the partition function of a single cusp is one-loop exact.

A separate motivation for including cusps (in supergravity) is discussed in section \ref{sec:super}.

\section{Universal origin of accumulation points}\label{sec:renormalization_proposal}
There is a nice fact about three-manifolds that makes it possible to resolve all accumulation points at once. Basically, accumulation points have a universal origin.\footnote{This follows from Jorgensen's theorem presented as theorem 5.12.1 of \cite{thurston2022geometry}, see also \cite{gromov1981hyperbolic}.
} In each case, what is happening is the formation of an infinite cusp,
\be
\begin{tikzpicture}[scale=0.35, rotate=0, baseline={([yshift=-0.1cm]current bounding box.center)}]
\node at (22cm,-8cm) {\includegraphics[scale = .5]{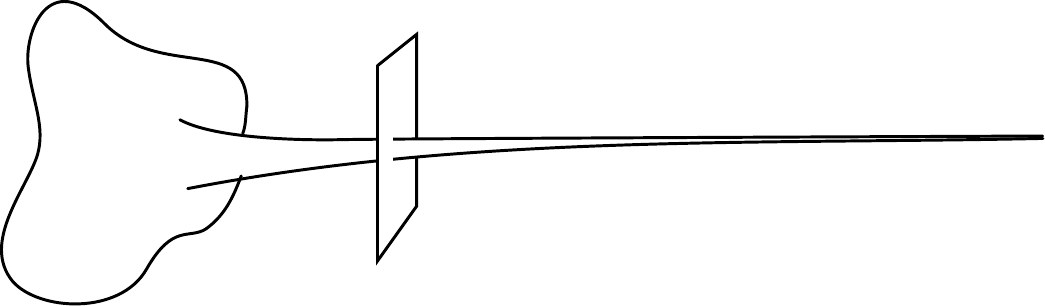}};
\node at (36.8cm,-7.65cm) {$\rightarrow \infty$};
%\node[align=center] at (24cm,-12cm) {cross section is a torus\\with modular parameter $\tau$};
\end{tikzpicture}
\ee
The cross section that one gets by cutting through the cusp with the plane shown above is conformal to a torus with some modular parameter $\tau$:
\be
\begin{tikzpicture}[scale=1.75, rotate=0, baseline={([yshift=-0.1cm]current bounding box.center)}]
  % Axes
  \draw[->] (-0.2,0) -- (1.5,0) node[right] {Re};
  \draw[->] (0,-0.2) -- (0,1.1) node[above] {Im};
  % Fundamental parallelogram
  \filldraw[thick, lightgray]
    (0,0) -- (1,0) -- ++(0.6,1) -- (0.6,1) -- cycle;
  \draw[thick]
    (0,0) -- (1,0) -- ++(0.6,1) -- (0.6,1) -- cycle;
  % Vertex labels
  \draw[thick] (.3-.05,.5) -- (.3+.05,.5);
  \draw[thick] (1+.3-.05,.5) -- (1+.3+.05,.5);
  \draw[thick] (1.1,1+.05) -- (1.1,1-.05);
  \draw[thick] (1.1-.05,1+.05) -- (1.1-.05,1-.05);
  \draw[thick] (.5,-.05) -- (.5,.05);
  \draw[thick] (.5-.05,-.05) -- (.5-.05,.05);
  \node[fill, circle, inner sep=1pt, label=below left:$0$] at (0,0) {};
  \node[fill, circle, inner sep=1pt, label=below:$1$] at (1,0) {};
  \node[fill, circle, inner sep=1pt, label=above:$\tau$] at (0.6,1) {};
\end{tikzpicture}
\ee
This cusped manifold is the limit of an accumulating set of smooth hyperbolic manifolds, constructed by cutting the cusp at some location and gluing in a solid torus in such a way that the cycle $c\tau + d$ becomes contractible. Here $c,d$ should be relatively prime integers, and Thurston's hyperbolic Dehn surgery theorem (theorem 5.8.2 of \cite{thurston2022geometry}) shows that for all but a finite number of pairs $\{c,d\}$, this topology admits a hyperbolic metric.

When $c,d$ are large, the resulting manifold $M_{c,d}$ resembles the original cusped manifold out to a distance $\log|c\tau + d|$ ``along the cusp,'' after which point the $c\tau + d$ cycle smoothly contracts.
% \be
% \begin{tikzpicture}[scale=0.35, rotate=0, baseline={([yshift=-0.1cm]current bounding box.center)}]
% \node at (12cm,0) {\includegraphics[scale = .5]{figures/2.pdf}};
% \node[align=center, gray] at (24cm,2cm) {$c\tau + d$ contracts};
% \draw[thick, ->, gray] (19cm,1.5cm) -- (16.5cm,.5cm);
% \filldraw[white] (13.15,-1.5) circle (5pt);
% \draw[thick,|-|, gray] (12,-1.5) -- (16.2,-1.5);
% \draw[thick, ->, gray] (17,-3.1) -- (14.1,-2);
% \node[gray] at (21,-3) {$\log|c\tau+d|$};
% \end{tikzpicture}
% \ee
The volume approaches that of the original cusped manifold from below \cite{neumann1985volumes}
\be
\text{vol}(M_{c,d}) \approx  \text{vol}(M_{\infty}) - \pi^2\frac{\text{Im}(\tau)}{|c\tau+d|^2}.
\ee

For the purposes of quantum gravity, these accumulation points are bad because they make the sum over the saddle points $M_{c,d}$ divergent.\footnote{Based on what we've said so far, this is actually not completely clear. It is clear that the suppression factor $e^{-\text{vol}(M)/(4\pi G)}$ coming from the Einstein-Hilbert action will not make the sum converge, since the volume approaches a limiting value as $c,d$ go to infinity. But one could imagine that the integral over small fluctuations around the saddle point could give a correction that makes the sum converge. Unfortunately, the one-loop determinant does not decay rapidly enough \cite{Maloney:2007ud,Giombi:2008vd} as we will review below.} This point was discussed by Collier, Eberhardt and Zhang \cite{Collier:2023fwi}, who pointed out that precisely this divergent sum was already encountered in quantum gravity work \cite{Maldacena:1998bw,Dijkgraaf:2000fq,Maloney:2007ud}. Maloney and Witten assigned the sum a finite value using zeta function regularization \cite{Maloney:2007ud}, implicitly subtracting an infinite contribution. In this paper we treat the sum using a related procedure in which the divergence is made explicit. We then include a new contribution from the limiting cusp, as a counterterm that cancels the divergence.

\section{Cancellation in a simple case} 
In this section, we show that a single cusp can be used to cancel the divergence in the simplest setting in which an accumulation point arises: the sum over so-called SL$(2,\mathbb{Z})$ black holes. The procedure is as follows:
\begin{enumerate}
  \item Introduce a convergence factor into the sum. To preserve modular invariance, this factor should depend on an intrinsic feature of the geometry, and not the modular frame in which we consider it. A convenient choice is to insert $e^{-\epsilon / \ell_{c,d}}$ where $\ell_{c,d} = 2\pi \text{Im}(\tau)/|c\tau+d|^2$ is the length of the closed geodesic where $c\tau+d$ contracts, as measured by the metric of the classical solution.
  \item Include a new contribution from the limiting cusp, with a coefficient $\mu < 0$.
  \item Take $\epsilon \to 0$, adjusting $\mu(\epsilon)$ so that the cusp cancels the divergence in the sum.
\end{enumerate}
We emphasize that according to the simplest rules of quantum gravity, the limiting cusp would not have contributed to (\ref{qg:label}). This is because it is not a closed manifold -- it contains a single point of asymptotic boundary. But the rules can be modified to allow cusps, with an arbitrary ``fugacity'' coefficient $\mu$ per cusp.\footnote{This freedom has been used before in two-dimensional gravity. In section 5.5 of \cite{Stanford:2019vob} a cusp with Ramond spin structure was included, and the fugacity was related to a property of the dual matrix integral (the number of rows minus columns for the supercharge). In section 2.2 of \cite{Gao:2021uro} a ``cusp-like counterterm'' was included to remove the divergence associated to small loops of end of the world branes.} The nontrivial thing we will check below is that the same $\mu(\epsilon)$ will work for accumulation points with arbitrary $\tau$.

% To handle the more general case with multiple incipient cusps, one can modify the sum over hyperbolic manifolds (\ref{qg:label}) to
% \be
% \sum_{\text{topology}}\int_{\text{geometry}} \hspace{-10pt}\exp\left(-I\right) \ \ \longrightarrow \ \ \ 
% \lim_{\epsilon\to 0} \sum_{\text{topology}}\mu(\epsilon)^{\#\text{ cusps}}\exp\left(-\epsilon\hspace{-5pt}\sum_{\substack{\text{short} \\ \text{geodesics}}}\hspace{-5pt}\frac{1}{\ell}\right) \int_{\text{geometry}} \hspace{-10pt}\exp\left(-I\right).
% \ee
% The RHS includes manifolds with cusps, and it also includes a convergence factor involving the lengths of short geodesics.\footnote{A three dimensional hyperbolic manifold can have an infinite number of primitive closed geodesics. But it can only have a finite number of short geodesics, less than a universal constant referred to in \cite{gromov1981hyperbolic} as the Kazhdan-Margulis constant. These short geodesics all sit at the ends of would-be cusps that have been closed off into ``tubes'' by gluing in a small solid torus.} For $\epsilon > 0$, the sum over hyperbolic manifolds with volume less than any finite $V$ will be finite. If we choose $\mu(\epsilon)$ to cancel the divergence of any single accumulation point, the full sum will also be finite as $\epsilon \to 0$ for any $V$. This is because the potential divergences come from very long would-be cusps, and the gravity path integral is local enough to reduce this analysis to the case of a single isolated cusp.

\subsection{The SL\texorpdfstring{$(2,\mathbb{Z})$}{(2,Z)} black holes} 
It is easiest to describe the accumulating set if we first describe the cusp manifold that they accumulate to. This is the geometry $\mathbb{H}^3/\mathbb{Z}\times \mathbb{Z}$ (the ``doubly infinite cusp'' in \cite{gromov1981hyperbolic}):
\begin{align}\label{cuspmetric:label}
\d s^2 = \d\rho^2+  e^{2\rho}(\d x^2 + \d y^2), \hspace{20pt} x + \i y &\sim x + \i y +2\pi \\ &\sim x + \i y + 2\pi\tau.\notag
\end{align}
At $\rho\to \infty$ there is an asymptotic boundary conformal to a torus, and at $\rho \to -\infty$ there is the cusp itself. For any fixed $\rho$, the cross section is a flat torus with modular parameter $\tau$.

The accumulating set is obtained by the procedure described in section \ref{sec:renormalization_proposal}: topologically we cut at some value of $\rho$ and replace the $\rho \to -\infty$ region with a solid torus. If the solid torus is chosen so that e.g.~the cycle parametrized by $x$ becomes contractible, the explicit metric is
\begin{align}
\d s^2 = \d\rho^2+  \sinh^2(\rho)\d x^2 + \cosh^2(\rho)\d y^2, \hspace{20pt} x + \i y &\sim x + \i y +2\pi \\ &\sim x + \i y + 2\pi\tau.\notag
\end{align}
Now $\rho$ runs from zero (where the $x$ cycle contracts) to infinity (the asymptotic boundary). The other solutions are all equivalent to this after making a modular transformation of the boundary torus \cite{Maldacena:1998bw,Dijkgraaf:2000fq}. We review these modular transformations in appendix \ref{app:modular_transformations_of_the_torus}. Explicitly,
\begin{align}
\d s^2 = \d\rho^2+  \sinh^2(\rho)\d X^2 + \cosh^2(\rho)\d Y^2, \hspace{20pt} X + \i Y &\sim X + \i Y +2\pi \\ &\sim X + \i Y + 2\pi\gamma\tau,\notag
\end{align}
where
\be
\gamma\tau = \frac{a\tau + b}{c\tau+d}, \hspace{20pt} ad-bc = 1, \hspace{20pt}a,b,c,d \in \mathbb{Z}.
\ee
It is conventional to write $\tau = \tau_1 + \i \tau_2$, with $\tau_1,\tau_2\in \mathbb{R}$. Then
\begin{align}\label{dep}
(\gamma\tau)_1 &= \frac{a}{c} - \frac{c\tau_1 + d}{c|c\tau+d|^2}\\
(\gamma\tau)_2 &= \frac{\tau_2}{|c\tau+d|^2}.\label{depona}
\end{align}
Here are some comments about this geometry and its contribution to (\ref{qg:label}).
\begin{enumerate}
  \item
One can also write this metric in terms of the original coordinates $x,y$. It is useful to define $e^{\sigma} = e^{\rho}/(2|c\tau+d|)$. Then
\begin{align}
\d s^2
=\d\sigma^2 &+ e^{2\sigma}(\d x^2 + \d y^2) \\ &+ \left(\text{terms that become important for $e^{2\sigma}\sim |c\tau+d|^{-2}$}\right).
\end{align}
This formula makes it clear that for fixed $\tau$, if we make $c,d$ large, the metric will resemble the cusp (\ref{cuspmetric:label}) for a length that grows as $\log|c\tau+d|$.

\item The length of the closed geodesic at $\rho = 0$ is
\be
\text{length of closed geodesic} = 2\pi (\gamma\tau)_2 = 2\pi \frac{\tau_2}{|c\tau+d|^2}.
\ee

\item The regularized volume is
\be
\text{vol} = -\pi^2(\gamma\tau)_2.
\ee

\item From the perspective of (\ref{qg:label}), these manifolds are saddle points, with saddle point values of the action given by (\ref{EHaction:label}), and therefore weighting
\begin{align}
e^{-I} &= \exp\left(\frac{\pi}{4G}\frac{\tau_2}{|c\tau+d|^2}\right)\\
&=|q\bar{q}|^{-k}\big|_\gamma.
\end{align}
In the last line we took the opportunity to introduce some convenient notation
\be
q = e^{2\pi \i \tau}, \hspace{20pt} \hspace{20pt} k = \frac{1}{16 G}.
\ee
Also $\bar{q} = e^{-2\pi\i\bar{\tau}}$ is the complex conjugate of $q$, and the notation $f(\tau)|_\gamma$ means $f(\gamma \tau)$.

\item Associated to each of these saddle points is a family of geometries with the same topology and boundary conditions. The integral over these geometries was evaluated in \cite{Maloney:2007ud,Giombi:2008vd}.\footnote{See \cite{Bytsenko:1997ru,Yin:2007gv} for related earlier work.}, with the result
\be\label{mw:label}
\int_{\text{geometry}}e^{-I} = \left(|\bar{q} q|^{-k + \frac{1}{24}}\frac{|1-q|^2}{|\eta(\tau)|^2}\right)_{\gamma}
\ee
where $\eta$ is the Dedekind $\eta$ function. The entire expression on the RHS is simply the Virasoro vacuum character, transformed by $\gamma$.

\item In principle, one should sum over all distinct topologies (relative to a fixed choice of cycles on the asymptotic boundary torus). This means that we sum over distinct choices of primitive contractible cycle $\Omega_1$, which means to sum over relatively prime $c,d$ up to the equivalence $(c,d)\sim (-c,-d)$ \cite{Maloney:2007ud}
\begin{align}
Z(\tau) &= \sum_{\text{topology }(c,d)}\int_\text{geometry}e^{-I} \\ &=\frac{1}{\sqrt{\tau_2}|\eta(\tau)|^2}\sum_{(c,d)= 1}\left(\sqrt{\tau_2}|\bar{q} q|^{-k + \frac{1}{24}}|1-q|^2\right)_{\gamma}\label{mwanswer}
\end{align}
The last line is just a slightly rewritten version of (\ref{mw:label}), where a modular invariant factor $\sqrt{\tau_2}|\eta(\tau)|^2$ was pulled outside the action of $\gamma$ and therefore the sum. To take into account $\{c,d\}\to -\{c,d\}$, one could put a factor of one half out front and then sum over both positive and negative relatively prime $c,d$. Alternatively, when $c$ is nonzero one can require it to be positive, and when $c$ is zero one can require $d$ to be positive.

\item The real part of $\gamma\tau$ depends on $a$, but the sum is over $c$ and $d$, with no explicit mention of $a$. This makes sense because we can regard $a$ as a function of $c,d$ by using the extended Euclidean algorithm to find $a,b$ such that $ad-bc = 1$. The value of $a$ is not unique, but its non-uniqueness consists of adding an integer multiple of $c$, which does not affect (\ref{mwanswer}). This non-uniqueness corresponds to adding a multiple of the contractible cycle $\Omega_1$ to $\Omega_2$. This does not change the manifold.

\item The sum is divergent \cite{Dijkgraaf:2000fq,Maloney:2007ud}. For large $c,d$ it looks like
\be
\sum_{(c,d)=1} \frac{\text{something}}{|c\tau+d|} \sim \int \frac{\d^2 x}{|x|}.
\ee
\end{enumerate}

\subsection{Renormalization of the sum over SL\texorpdfstring{$(2,\mathbb{Z})$}{(2,Z)} black holes}
In \cite{Maloney:2007ud}, Maloney and Witten define the sum (\ref{mwanswer}) as the analytic continuation to $s = 1/2$ of a generalized sum that converges for $s >1$:
\be\label{zeta:label}
Z(\tau) = \mathop{\mathrm{continuation}}_{s\to 1/2}\frac{1}{\sqrt{\tau_2}|\eta(\tau)|^2}\sum_{(c,d)=1}\left(\tau_2^s|\bar{q} q|^{-k + \frac{1}{24}}|1-q|^2\right)_{\gamma}.
\ee
We propose a different prescription that leads to the same answer:
\be\label{regularizedMW}
Z(\tau) = \frac{1}{\sqrt{\tau_2}|\eta(\tau)|^2}\lim_{\epsilon\to 0} \left[- \frac{6}{\sqrt{\pi}\sqrt{\epsilon}}+\sum_{(c,d)=1}\left(\sqrt{\tau_2}|\bar{q} q|^{-k + \frac{1}{24}}|1-q|^2e^{-\epsilon/\tau_2}\right)_{\gamma} \right].
\ee
The convergence factor $e^{-\epsilon/\tau_2}$ weights geometries by $e^{-2\pi\epsilon / \ell}$ where $\ell$ is the length of the closed geodesic. The two expressions (\ref{zeta:label}) and (\ref{regularizedMW}) are equal by the general agreement between zeta function regularization and renormalization with a smooth cutoff.\footnote{See \href{https://terrytao.wordpress.com/2010/04/10/the-euler-maclaurin-formula-bernoulli-numbers-the-zeta-function-and-real-variable-analytic-continuation}{this blog post} by Terence Tao.}. See also appendix \ref{app:zeta_regularization}.

The Maloney-Witten regularization (\ref{zeta:label}) is a form of zeta function regularization, and it has the mixed blessing of obscuring both the divergence and the counterterm that needs to be added in order to cancel it. Having exposed this divergence with the alternative formula (\ref{regularizedMW}), we would like to find a geometrical interpretation for the necessary $-6/\sqrt{\pi\epsilon}$ counterterm.

Our proposal is that this can be provided by the contribution of the limiting cusp itself (\ref{cuspmetric:label}). For this to work, we would like the contribution of the cusp geometry to be proportional to $(\sqrt{\tau_2}|\eta(\tau)|^2)^{-1}$. If this is true, then by adjusting the fugacity correctly we will be able to cancel the divergence in (\ref{regularizedMW}). We find that this is true in a one-loop approximation to the gravitational path integral, although it is a little bit subtle: the one-loop determinant diverges and this is the answer we get after making a modular invariant regularization of that determinant.

One can start by computing the action (\ref{EHaction:label}) for the cusp metric (\ref{cuspmetric:label}). This turns out to simply be zero. Next we study the one-loop determinant. Because the cusp is $\mathbb{H}^3/\mathbb{Z}\times \mathbb{Z}$, one can do this using a sum over images, following \cite{Giombi:2008vd}. To explain the procedure, we consider the simpler case of a massive scalar field:\footnote{This case is somewhat formal because of a UV divergence in the $t$ integral not present for the graviton.}
\begin{align}
\log Z_{\text{1-loop}} &= -\frac{1}{2}\log\text{det}(-\Delta+M^2)\\ &= \frac{1}{2}\int_0^\infty \frac{\d t}{t}\int\d^3\x \sqrt{g} K^{\mathbb{H}^3/(\mathbb{Z}\times\mathbb{Z})}(t,\x,\x)\\
&= \frac{1}{2}\text{vol}(\mathbb{H}^3/(\mathbb{Z}\times\mathbb{Z}))\int_0^\infty \frac{\d t}{t}K^{\mathbb{H}^3}(t,0,0) + \frac{1}{2}\sum_{(n,m)\neq (0,0)}\int_0^\infty \frac{\d t}{t} \int_{\mathbb{H}^3/(\mathbb{Z}\times\mathbb{Z})} \hspace{-30pt}\d^3 \x \sqrt{g} K^{\mathbb{H}^3}(t,r(\x,\gamma_{n,m}\x))\notag
\end{align}
Here $K(t,\x,\x') = \langle \x|e^{(\Delta-M^2)t}|\x'\rangle$ is the heat kernel, see \cite{Vassilevich:2003xt}. The first term in the final line is canceled by a bulk counterterm and we will ignore it from now on. The quantity $r$ in the second term is the geodesic distance between points in the covering space that are related by the $n,m$ action within $\mathbb{Z}\times \mathbb{Z}$:
\be
\gamma_{n,m}\{x+\i y,\rho\} = \{x+\i y + 2\pi(n + m\tau), \rho\}.
\ee
The geodesic distance between a point and its $\gamma_{n,m}$ image in the covering hyperbolic space is
\be
\cosh(r(\x,\gamma_{n,m}\x)) = 1 + \frac{(2\pi)^2e^{2\rho}}{2}|n+m\tau|^2.
\ee
This formula allows us to change variables from $\rho$ to $r$. The integration measure is
\be
\d^3\x \sqrt{g}= e^{2\rho}\d\rho \d x \d y = \frac{\sinh(r)}{(2\pi)^2|n+m\tau|^2}\d r \d x \d y
\ee
So 
\begin{align}
\log Z_{\text{1-loop}} &= \frac{1}{2}\sum_{(n,m)\neq (0,0)}\int_0^\infty \frac{\d t}{t} \int_0^\infty \d r\frac{(2\pi)(2\pi\tau_2)\sinh(r)}{(2\pi)^2|n + m\tau|^2} K^{\mathbb{H}^3}(t,r)\\
&=\frac{1}{2}\sum_{(n,m)\neq (0,0)}\frac{\tau_2}{|n+m\tau|^2}\int_0^\infty \frac{\d t}{t} \frac{e^{-(1+M^2)t}}{4\pi^{3/2}t^{1/2}},\label{analogous}
\end{align}
where we used the explicit formula for the scalar heat kernel
\be
K^{\mathbb{H}^3}(t,r) = \frac{1}{(4\pi t)^{3/2}}\frac{r}{\sinh(r)}e^{-(1+M^2)t-\frac{r^2}{4t}}.
\ee

For the case of the graviton determinant, the structure is similar, but the heat kernel $K$ is more complicated. Fortunately, the results from \cite{Giombi:2008vd} can be adapted to our case by setting $\tau = 0$ in (4.25) and (4.26) of \cite{Giombi:2008vd}.\footnote{We checked this by doing the $r$ integral numerically. For spin one gauge fields the analogous statement can be checked by hand.} The analog of (\ref{analogous}) is
\begin{align}
\log Z_{\text{1 loop}} &= \sum_{(n,m)\neq (0,0)}\frac{\tau_2}{|n+m\tau|^2}\int_0^\infty\frac{\d t}{t}\frac{e^{-t} - e^{-4t}}{4\pi^{3/2}\sqrt{t}}\\ &= \frac{1}{2\pi}\sum_{(n,m)\neq (0,0)}\frac{ \tau_2}{|n+m\tau|^2}.\label{cusponeloop:label}
\end{align}
This sum is logarithmically divergent, meaning that the partition function on the cusp diverges. However, if we regularize it in a modular invariant way, the divergence is independent of $\tau$ and can be absorbed into $\mu$. One modular invariant way to regularize the sum is (see appendix \ref{app:eisenstein})
\be\label{regularized}
\frac{1}{2\pi}\sum_{(n,m)\neq (0,0)}\frac{ \tau_2}{|n+m\tau|^2}e^{-\varepsilon |n+m\tau|^2/\tau_2} = \frac{\log\frac{1}{\varepsilon}}{2} + \frac{\gamma}{2} - \log(2) - \log\left[\sqrt{\tau_2}|\eta(\tau)|^2\right] + O(\varepsilon).
\ee
Here we included a smoothed out version of a cutoff $|n+m\tau| < \sqrt{\tau_2/\epsilon}$ on the image sum. \footnote{Since the geodesic distance between image points depends only on $|n+m\tau|e^\rho$, we could get the same effect by smoothly cutting off $\rho$ so that $e^\rho > \sqrt{\epsilon/\tau_2}$. This has the effect of cutting off the cusp so that the cross sectional area of the torus is at least $\epsilon$ -- this is clearly modular invariant.} So at one loop
\be
Z_{\text{cusp}}(\tau) = \mu \cdot \frac{e^{\gamma/2}}{2\sqrt{\varepsilon}}\cdot  \frac{1}{\sqrt{\tau_2}|\eta(\tau)|^2}
\ee
where $\mu$ is the arbitrary fugacity. After choosing $\mu \cdot \frac{e^{\gamma/2}}{2\sqrt{\varepsilon}} = -\frac{6}{\sqrt{\pi\epsilon}}$ the cusp will provide the needed counterterm in (\ref{regularizedMW}).

As a next step, one ought to study higher loops. We are assuming that they vanish (or can be absorbed into rescaling $\mu$) but this needs to be checked.

\section{Cancellation in general}

Ref.~\cite{Collier:2023fwi} pointed out that because of the locality of the gravity path integral and the universal origin of accumulation points, if one can make sense of the SL$(2,\mathbb{Z})$ black holes then one can treat the general case.

To make this argument, suppose we have a family of topologies with accumulating hyperbolic volumes.\footnote{In the volume spectrum there are accumulation points of accumulation points and so on. This sounds intimidating but these correspond to the formation of multiple cusps at different locations on the three manifold, and they can be handled separately.} This family consists of Dehn fillings of some torus within the three manifold. The path integral can be decomposed into the region outside the torus and whatever we fill it in with:
\be
Z = \langle \text{rest of manifold } | \,(c,d)\text{ Dehn filling}\rangle.
\ee
Here, the bra and ket are vectors in the Virasoro TQFT Hilbert space on the torus \cite{Collier:2023fwi}. The contribution of the limiting cusp would be
\be
\langle \text{rest of manifold } | \text{ cusp}\rangle.
\ee
For the case of the SL$(2,\mathbb{Z})$ black holes, the rest of the manifold is replaced by the state $\langle \tau,\bar\tau|$ corresponding to the boundary conditions for the asymptotic torus. These states span the Hilbert space on the torus, so if the cusp cancels the divergence from the Dehn fillings for all $\tau,\bar{\tau}$, then it implies the cancellation for a general manifold.

Note that the cusp state has a simple representation\footnote{Tom Hartman pointed out that this is proportional to the Virasoro Turaev-Viro state for a zero-extrinsic curvature torus boundary with a geodesic with zero dihedral angle \cite{Hartman:2025ula}.}
\be
|\text{cusp}\rangle = (\text{const.})\int_0^\infty \d P |P,P\rangle.
\ee
where $P,\bar{P}$ are so-called Liouville momenta, and $|P,\bar{P}\rangle$ satisfies
\be
\langle \tau,\bar\tau|P,\bar{P}\rangle = \frac{q^{P^2}}{\eta(\tau)}\cdot \frac{\bar{q}^{\bar P^2}}{\eta(\bar\tau)}.
\ee

\section{\texorpdfstring{$\mathcal{N}=1$}{N=1} supersymmetry}\label{sec:super}
One can also consider the contribution of a cusp in $\mathcal{N}=1$ supergravity. The one-loop determinant is a product
\be
Z_{\text{1 loop}} = Z_{\text{1 loop}}^{\text{graviton}}\cdot Z_{\text{1 loop}}^{\text{gravitino}}.
\ee
We already computed the one-loop determinant for the graviton. The one-loop determinant for the gravitino can be computed in a similar way, but now borrowing heat kernel results from (6.9) and (7.28) of \cite{David:2009xg}. One finds
\begin{align}
\log Z_{\text{1 loop}}^{\text{gravitino}} &= -\sum_{(n,m)\neq (0,0)} (-1)^{\sigma_n n + \sigma_m m}\frac{\tau_2}{|n+m\tau|^2}\int_0^\infty \frac{\d t}{t} \frac{e^{(\frac{9}{4} - \frac{5}{2})t} - e^{(-\frac{3}{4} - \frac{3}{2})t}}{4\pi^{3/2}\sqrt{t}}\\
&=-\frac{1}{2\pi}\sum_{(n,m)\neq (0,0)} (-1)^{\sigma_n n + \sigma_m m}\frac{\tau_2}{|n+m\tau|^2}
\end{align}
Here the sign factor $(-1)^{\sigma_n n + \sigma_m m}$ depends on the choice of spin structure on the torus. 

There is a particularly simple answer in the case of purely periodic boundary conditions (Ramond spin structure on both cycles of the torus). In this case, $\sigma_n = \sigma_m = 0$ and the graviton and gravitino one-loop determinants cancel exactly. We are left with a very simple partition function
\be
Z_{\text{1 loop}} = 1 \hspace{20pt} (\text{periodic on both cycles}).
\ee
This answer is consistent with the fact that the purely periodic spin structure corresponds to an index $\tr_{R}\left[(-1)^F\dots\right]$ that is independent of $\tau$. Naively, the bulk answer for this index should just be zero, because there is no smooth three-manifold that can fill in torus boundary with purely periodic spin structure. But we see that in fact one can make the index nonzero by including a purely Ramond cusp with a nonzero fugacity. This is exactly parallel to the situation in $\mathcal{N} = 1$ super JT gravity, see section 5.5 of \cite{Stanford:2019vob}. We regard this as further evidence that it is reasonable to include a nonzero fugacity for cusps.

There are three other spin structures to analyze, where we include antiperiodic boundary conditions on one or both of the cycles. These three cases are related by modular transformations, so we can consider just one of them, and we will take the case $\sigma_n = 1,\sigma_m = 0$ corresponding to $\tr_{NS}[(-1)^F\dots]$:
\begin{align}
\log Z_{\text{1 loop}} &= \frac{1}{2\pi}\sum_{(n,m)\neq (0,0)}\frac{\tau_2}{|n+m\tau|^2}(1 - (-1)^n)e^{-\varepsilon |n+m\tau|^2/\tau_2}\\
&= \frac{2}{2\pi}\sum_{(n,m)\neq (0,0),n \text{ odd}}\frac{\tau_2}{|n+m\tau|^2}e^{-\varepsilon |n+m\tau|^2/\tau_2}\\
&=\frac{2}{2\pi}\sum_{(n,m)\neq (0,0)}\left(\frac{\tau_2}{|n+m\tau|^2}e^{-\varepsilon |n+m\tau|^2/\tau_2} - \frac{\tau_2}{|2n+m\tau|^2}e^{-\varepsilon |2n+m\tau|^2/\tau_2}\right)
\\
&=\frac{2}{2\pi}\sum_{(n,m)\neq (0,0)}\left(\frac{\tau_2}{|n+m\tau|^2}e^{-\varepsilon |n+m\tau|^2/\tau_2} - \frac{1}{2}\frac{\tau_2/2}{|n+m\tau/2|^2}e^{-2\varepsilon |n+m\tau/2|^2/(\tau_2/2)}\right)\\
&= \frac{\log\frac{1}{\varepsilon}}{2}+\frac{\gamma}{2}-\log(2) + \log \frac{|\eta(\frac{\tau}{2})|^2}{\sqrt{\tau_2}|\eta(\tau)|^4} + O(\varepsilon).
\end{align}
So for the spin structure that is periodic along the $\tau$ cycle and antiperiodic along the 1 cycle, we get (assuming one-loop exactness)
\be\label{cuspsuper:label}
Z_\text{cusp}(\tau) = \mu \cdot \frac{e^{\gamma/2}}{2\sqrt{\varepsilon}}\cdot  \frac{|\eta(\tau/2)|^2}{\sqrt{\tau_2}|\eta(\tau)|^4}.
\ee
For this spin structure, the sum over SL$(2,\mathbb{Z})$ black holes was computed by Maloney and Witten \cite{Maloney:2007ud}, see (7.16). This sum is divergent and was treated with zeta function-like regularization. As in the bosonic case, one could alternatively define the sum with a smooth modular invariant cutoff. The result would then be the Maloney-Witten answer plus a divergence that is proportional to (\ref{cuspsuper:label}). So we conclude that as for pure gravity, the cusp can be used to renormalize the sum over accumulation points in pure $\mathcal{N} = 1$ supergravity.

\section{Discussion}
In this paper we proposed to renormalize accumulation points in 3d gravity by cancelling the divergence against the cusped geometry that they accumulate to. 

An obvious loose end is the question of whether the cusp partition function is one-loop exact. Note that in \cite{Collier:2024mgv}, a different prescription is used for a cusp partition funcion, see appendix \ref{app:cusp_vs_extreme_conical_defect}. However this leads to an answer that agrees with ours after a sum over modular images.

An unsatisfying aspect of our proposal is that the convergence factor we used depends on the length $\ell_{c,d}$ in the classical solution. This is a nonlocal thing to insert into the path integral (\ref{qg:label}). This objection also applies to \cite{Maloney:2007ud}. It might be interesting to explore alternatives.

In this paper we have not addressed the contribution of off-shell manifolds, despite the fact that in the spin-zero sector they are expected to contribute at the same order (order $e^{0/G_N}$) as the cusp. In principle, the counterterm that is needed could be modified by the sum over off-shell manifolds. In fact, we expect that at least some off-shell manifolds resemble the cusp: this is the case in JT gravity where off-shell topologies are off-shell precisely becuase the trumpet ``wants'' to turn into a cusp (the saddle point at infinity with $b = 0$).

\section*{Acknowledgements} 

We are grateful to Yiming Chen, Thomas Hartman, Daniel Jafferis, and Henry Maxfield for important insights, and to Antoine Song and Zhenbin Yang for a discussion that motivated this work. This work was supported in part by DOE grants DE-SC0021085 and DE-SC0026143 and by a grant from the Simons foundation (926198, DS).

\appendix

\section{Modular transformations of the torus}\label{app:modular_transformations_of_the_torus}
Up to rescaling, a flat $\mathbb{T}^2$ can be regarded as the complex plane identified by a lattice generated by vectors
\be
\omega_1 = 2\pi, \hspace{20pt} \omega_2 = 2\pi\tau.
\ee
Exactly the same torus, but with a different fundamental domain in the complex plane, arises if we use $\Omega_1,\Omega_2$ related to $\omega_1,\omega_2$ by an SL$(2,\mathbb{Z})$ transformation:
\be
\left(\begin{array}{c}\Omega_2 \\ \Omega_1\end{array}\right) =\left(\begin{array}{cc} a & b \\ c  & d \end{array}\right)\left(\begin{array}{c}\omega_2 \\ \omega_1\end{array}\right), \hspace{20pt} ad - bc = 1, \ \  \{a,b,c,d\}\in \mathbb{Z}.
\ee
For example, \vspace{-25pt}
\be
\begin{tikzpicture}[scale=1, rotate=0, baseline={([yshift=-0.1cm]current bounding box.center)}]
  %— PARAMETERS —%
  \pgfmathsetmacro{\tauone}{0.6}   % \tau_1
  \pgfmathsetmacro{\tautwo}{1.2}   % \tau_2
  \pgfmathsetmacro{\a}{2}
  \pgfmathsetmacro{\b}{1}
  \pgfmathsetmacro{\c}{1}
  \pgfmathsetmacro{\d}{1}

  \node at (-5,1) {$\left(\begin{array}{cc} a & b \\ c  & d \end{array}\right)=\left(\begin{array}{cc} 2 & 1 \\ 1  & 1 \end{array}\right)$};
  \node at (-1.75,1) {$\rightarrow$};

  %— DEFINE BASIS VECTORS —%
  \coordinate (O)  at (0,0);
  \coordinate (w1) at (1,0);                             % ω₁
  \coordinate (w2) at (\tauone,\tautwo);                 % ω₂
  \coordinate (W1) at ({\d*1+\c*\tauone},{\c*\tautwo});  % Ω₁ = d·ω₁ + c·ω₂
  \coordinate (W2) at ({\b*1+\a*\tauone},{\a*\tautwo});  % Ω₂ = b·ω₁ + a·ω₂

  %— AXES —%
  \draw[->] (-0.1,0) -- (1.8,0);% node[right] {$x$};
  \draw[->] (0,-0.1) -- (0,1.8);% node[above] {$y$};

  %— DRAW FUNDAMENTAL PARALLELOGRAMS —%
  % original
  \draw[thick,gray,dashed]
    (O) -- (w1) -- ($(w1)+(w2)$) -- (w2) -- cycle;
  % transformed
  \draw[thick,black,dashed]
    (O) -- (W1) -- ($(W1)+(W2)$) -- (W2) -- cycle;

  \fill[white] (W1) circle (4pt);
  \fill[white] (W2) circle (1pt);

  %— DRAW BASIS VECTORS —%
  \draw[->,gray,very thick] (O) -- (w1) node[below] {$\omega_1$};
  \draw[->,gray,very thick] (O) -- (w2) node[above]  {$\omega_2$};
  \draw[->,black,very thick]  (O) -- (W1) node[below right]       {$\Omega_1 = \omega_1+\omega_2$};
  \draw[->,black,very thick]  (O) -- (W2) node[above left]       {$\Omega_2 = \omega_1 + 2\omega_2$};
\end{tikzpicture}
\ee

So one can either think of the torus as the complex plane with identifications
\be\label{firstiden:label}
x+\i y \sim 2\pi, \hspace{20pt} x + \i y \sim x + \i y +2\pi \tau
\ee
or with identifications
\be\label{secondiden:label}
x+\i y \sim 2\pi(c\tau + d), \hspace{20pt} x + \i y \sim x + \i y +2\pi (a\tau+b).
\ee
One can make these two look more similar by defining new coordinates
\be
X + \i Y = \frac{x + \i y}{c\tau+d}
\ee
so that the second version of the identifications (\ref{secondiden:label}) becomes
\be\label{thirdiden:label}
X + \i Y \sim X + \i Y + 2\pi, \hspace{20pt} X + \i Y \sim X + \i Y +2\pi \gamma \tau.
\ee

\section{Zeta regularization \texorpdfstring{$=$}{=} smooth cutoff}\label{app:zeta_regularization}
One can simplify the job of comparing (\ref{zeta:label}) and (\ref{regularizedMW}) by subtracting convergent sums from both expressions. In particular, we can write $q\bar{q} = 1 + O(|c\tau+d|^{-2})$, and drop the correction because it is a convergent sum. Similarly, we can approximate
\begin{align}
|1-q|^2  &= 1 + q\bar{q} + q + \bar{q} \\
&\to  2 + q + \bar{q} \\ &\to 2.
\end{align}
The justification of dropping the terms linear in $q,\bar{q}$ is that the sum over these terms is convergent even in the limit $\epsilon \to 0$ or $s \to 1/2$. Roughly, the reason for this is that $q\approx e^{2\pi \i a/c}$ is approximately a random phase, leading to cancellations. More reliably, this is a $J = \pm 1$ term in the language of \cite{Keller:2014xba}, and sections 3.3 and 3.4 of that paper show convergence for $s \ge 1/2$.

Using these simplifications, showing (\ref{zeta:label})$=$(\ref{regularizedMW}) reduces to showing
\be\label{equi:label}
 \mathop{\mathrm{continuation}}_{s\to 1/2}
\sum_{(c,d)=1}\tau_2^s|_\gamma \stackrel{?}{=} \lim_{\epsilon \to 0}\left[- \frac{3}{\sqrt{\pi}\sqrt{\epsilon}}+\sum_{(c,d)=1}\sqrt{\tau_2}e^{-\epsilon/\tau_2}|_\gamma\right].
\ee
For general $s$, the sum on the LHS is known as the real analytic Eisenstein series, and for general $\epsilon$ the sum on the RHS is a related function that we will call $F$:
\begin{align}
E(\tau,s) &= \sum_{(c,d)=1}\tau_2^s|_{\gamma}\\
F(\tau,\epsilon) &= \sum_{(c,d)=1} \sqrt{\tau_2}e^{-\epsilon/\tau_2}|_\gamma.
\end{align}
They are related by Mellin and inverse Mellin transforms. In particular
\begin{align}
%E(s) &= \frac{1}{\Gamma(s-\frac{1}{2})}\int_0^\infty \d t K(t) t^{s-\frac{3}{2}}\\
F(\tau,\epsilon) &= \frac{1}{2\pi\i}\int_{C-\i\infty}^{C+\i\infty}\d s \ \epsilon^{\frac{1}{2}-s}\Gamma(s-\tfrac{1}{2})E(\tau,s).
\end{align}
The defining contour for the $s$ integral is to the right of all singularities in the integrand. $E(\tau,s)$ has a pole with residue $3/\pi$ at $s = 1$, and all further poles are to the left of $s = 1/2$. So, we should start out with a contour to the right of $s = 1$. Suppose we deform the contour leftwards, to just a bit left of $s = 1/2$. In the process we pick up the pole of $E$ at $s = 1$ and the pole of the $\Gamma$ function at $s = 1/2$. What remains is an integral over $s$ that lies entirely to the left of $s = 1/2$ and therefore vanishes as $\epsilon \to 0$. So
\be\label{B8:label}
F(\tau,\epsilon) = \frac{3}{\sqrt{\pi \epsilon}} + E(\tau,1/2) + (\text{vanishes as $\epsilon\to 0$}),
\ee
which implies (\ref{equi:label}). Note that $E(\tau,1/2) = 0$.

\section{Eisenstein series near \texorpdfstring{$s = 1$}{s=1}}\label{app:eisenstein}
The formula on the RHS of (\ref{regularized}) can be derived by writing
\be
\sum_{(n,m)\neq (0,0)}\left(\frac{ \tau_2}{|n+m\tau|^2}\right)^s = 2\zeta(2s)E(\tau,s)
\ee
and then processing the Fourier expansion of $E(\tau,s)$\footnote{See \href{https://en.wikipedia.org/wiki/Real_analytic_Eisenstein_series\#Fourier_expansion}{the Wikipedia page on Eisenstein series}.}:
\begin{align}
\frac{2\zeta(2s)E(\tau,s)}{2\pi} &= \frac{1}{2(s-1)} + \gamma-\log(2)+ \frac{\pi}{6}\tau_2 -\frac{\log(\tau_2)}{2}+\sum_{m = 1}^\infty \left(q^m + \bar{q}^m\right)\sigma_{-1}(m) + O(1-s)\label{eis2}
\end{align}
The sum involving the divisor function can be simplified:
\be
\sum_{m = 1}^\infty \left(q^m + \bar{q}^m\right)\sigma_{-1}(m) = \sum_{n=1}^\infty \frac{1}{n}\left(\frac{q^n}{1-q^n} + \frac{\bar{q}^n}{1-\bar{q}^n}\right) = -\sum_{m=1}^\infty \log|1-q^m|^2.
\ee
This then combines with the $\pi\tau_2/6$ to give $-\log|\eta(\tau)|^2$. Inserting into (\ref{eis2}) and then doing an inverse Mellin transform gives (\ref{regularized}). 

%%%D%%%
\section{Cusp vs.~extreme conical defect}\label{app:cusp_vs_extreme_conical_defect}
For some purposes, one can regard the cusp geometry as a limit of conical defect geometries. In particular, imagine cutting the cusp at some point and gluing in a solid torus with a conical deficit running along the center of the solid torus. In the limit that the total angle around this deficit vanishes, we recover the cusp geometry. Explicitly, for the simplest cusp with an asymptotic boundary, the metric would be
  \begin{align}\label{defectcusp:label}
\d s^2 = \d \rho^2 + \frac{\alpha^2}{(2\pi)^2}\left[\sinh^2(\rho)\d x^2 + \cosh^2(\rho) \d y^2\right], \hspace{20pt} x + \i y &\sim x + \i y +2\pi \\ &\sim x + \i y + 2\pi\tau.\notag
  \end{align}
For $\alpha = 2\pi$ this is a smooth manifold, for $0<\alpha<2\pi$ it has a conical deficit along the locus $\rho = 0$, and as $\alpha\to 0$ it approaches the cusp. 

The graviton one-loop determinant for conical defects $\alpha = 2\pi /N$ was studied in \cite{Benjamin:2020mfz}. The result is finite in the limit $N\to\infty$, approaching
\be\label{defect:label}
Z_{N\to \infty}(\tau) = \frac{1}{|\eta(\tau)|^2}.
\ee
This is different from our answer for the cusp one loop determinant. Technically, the difference arises as follows. The limit $N\to\infty$ of the summand from (4.21) of \cite{Benjamin:2020mfz} agrees with our (\ref{cusponeloop:label}). However, the subleading terms in $1/N$ are multiplied by increasingly divergent sums over $n,m$, and the large $N$ limit of their actual sum does not agree with our (\ref{cusponeloop:label}). 

The cusp was also regarded as a limit of a conical defect ($P_0\to 0$) in section 4 of \cite{Collier:2024mgv}. There, the authors were considering the figure eight knot complement, with an extreme conical singularity along the longitude of the figure eight knot.

Notice that the answer (\ref{defect:label}) is not modular invariant -- the one-loop determinant remembers which cycle of the torus gets the $\sinh^2(\rho)$ factor in (\ref{defectcusp:label}), even in the limit $\alpha \to 0$. In the context of \cite{Collier:2024mgv}, this can be seen from the fact that the torsion (which is the same as the graviton one-loop determinant) for the figure eight knot complement depends on which cycle of the torus gets the $\sinh^2(\rho)$ factor. For the longitudinal filling in which the conical singularity is along the knot itself, and for the $S$-dual meridian filling, pages 123 and 124 of \cite{porti1997torsion} give (after taking the inverse and removing a phase that we do not understand)
\be\label{porti:label}
Z^{\text{longitude}}_{\text{1-loop}}(\alpha) = \frac{1}{\sqrt{(\tfrac{3}{2}-\cos(\alpha))(\cos(\alpha)+\tfrac{1}{2})}}, \hspace{20pt} Z^{\text{meridian}}_{\text{1-loop}}(\alpha) = \frac{1}{\sqrt{17-8\cos(\tfrac{\alpha}{2})}}.
\ee
The ratio of these expressions in the limit $\alpha\to 0$ can be predicted from (\ref{defect:label}). The failure of modular invariance is associated to the cusp itself, not to the rest of the manifold, and locality implies that it can depend on the rest of the manifold only through the shape of the limiting cusp. For the figure eight knot complement, $\tau = 2\sqrt{3}\i$, so we expect 
\be
\frac{Z^{\text{longitude}}_{\text{1-loop}}(0)}{Z^{\text{meridian}}_{\text{1-loop}}(0)} = \frac{Z_{N\to \infty}(\tau)}{Z_{N\to \infty}(-1/\tau)} = |\tau| = 2\sqrt{3}.
\ee
This agrees with (\ref{porti:label}).

If one does try to make a modular invariant partition function by summing (\ref{defect:label}) over modular transformations, one finds a multiple of our answer:
\begin{align}
\sum_{(c,d) = 1} \frac{1}{|\eta(\tau)|^2}\Big|_\gamma &= \frac{1}{\sqrt{\tau_2}|\eta(\tau)|^2}\sum_{(c,d) = 1} \sqrt{\tau_2}\Big|_\gamma\\
&\to\lim_{\epsilon \to 0}\frac{1}{\sqrt{\tau_2}|\eta(\tau)|^2}F(\tau,\epsilon)\\
&= \frac{3}{\sqrt{\pi\epsilon}}\frac{1}{\sqrt{\tau_2}|\eta(\tau)|^2}
\end{align}
where we used (\ref{B8:label}) and $E(\tau,1/2) = 0$.

%%%E%%%
\section{SL\texorpdfstring{$(2,\mathbb{Z})$}{(2,Z)} black hole analysis in the energy basis}

In this appendix, we look at the divergence of the sum over SL\texorpdfstring{$(2,\mathbb{Z})$}{(2,Z)} black hole in the energy basis. We show that the MWK density of states \cite{Benjamin:2020mfz} and the Poincaré sum of extreme conical defect have the same piece of divergence that is related to a pole of Eisenstein series. And in both cases the contribution of interest comes from the spin zero sector. 

The modular-crossing kernel is given by
\be
\chi_P(\gamma\cdot \tau)=\int_0^\infty dP'\,\K_{P'P}^{(r,d)}[\one]\chi_{P'}(\tau)\quad\quad
\mathbb{K}^{(r,d)}_{P' P}[\one] = \epsilon \sqrt{\frac{8}{r}}e^{\frac{2\pi i }{r}( a P^2 + d P'^2)}\cos\left(\frac{4\pi}{r}P P'\right)
\ee
then the Maloney-Witten density of states can be written as 
\be
\rho_{MWK}(P,\bar{P})=\sum_{(c,d)=1}\K^{(c,d)}_{P\one}[\one] \bar{\K}^{(c,d)}_{\bar{P}\one}[\one]
\ee
This was rewritten in Appendix A of \cite{Benjamin:2020mfz} as
\begin{multline}
\rho_{MWK}(P,\bar{P})=\sum_{\ell=-\infty}^\infty\sum_{c=1}^\infty\frac{8}{c}\delta(j-\ell)\left[S(j,0;c)\cosh\frac{2\pi QP}{c}\cosh\frac{2\pi Q\bar{P}}{c}-S(j,-1;c)\cosh\frac{2\pi QP}{c}\cosh\frac{2\pi \tilde{Q}\bar{P}}{c}\right.\\
\left.-S(j,1;c)\cosh\frac{2\pi \tilde{Q}P}{c}\cosh\frac{2\pi Q\bar{P}}{c}+S(j,0;c)\cosh\frac{2\pi \tilde{Q}P}{c}\cosh\frac{2\pi \tilde{Q}\bar{P}}{c}\right]\label{mwkdensity}
\end{multline}
Here $j=P^2-\bar{P}^2$ is the spin and $S$ is the Kloosterman sum
\be
S(j,J;c)=\sum_{0\leq d<c, (c,d)=1}e^{2\pi i\frac{dj+(d^{-1})_cJ}{c}}
\ee
which in the special case $J=0$ reduces to the Ramanujan sum. The background charges are $Q=b+b^{-1}=\sqrt{\frac{c-1}{6}}$ and $\tilde{Q}=b-b^{-1}=\sqrt{\frac{c-25}{6}}$. The divergence piece of (\ref{mwkdensity}) is the same as the divergent piece of 
\be
\sum_{\ell=-\infty}^\infty\sum_{c=1}^\infty 8\delta(j-\ell)\left[2\frac{S(j,0;c)}{c}-\frac{S(j,-1;c)}{c}-\frac{S(j,1;c)}{c}\right]
\ee
because when $c$ is very large, we can ignore the $\cosh$ factors\footnote{$\sum_{c=1}^\infty\frac{S(j,J;c)}{c}$ is a divergent sum studied a lot by mathematicians. For more details see chapter 16 of \cite{iwaniec2021analytic}.}.
The sum can be regularized using zeta-function regularization. That was done in \cite{Benjamin:2020mfz} by introducing the regularized modular-crossing kernel.
\be
\K^{(s,\gamma)}_{P'P}[\one]=4\epsilon(s,\gamma)\left(\frac{2\pi}{c}\right)^s\frac{P^{2s-1}}{\Gamma(s)}e^{\frac{2\pi i }{c}( a P^2 + d P'^2)}\phantom{a}_0F_1(s;-\frac{4\pi^2}{c^2}P^2P'^2)
\ee
and then taking $s\rightarrow\frac{1}{2}$ limit. Here $\epsilon(s,\gamma)$ is an unimportant phase factor. After zeta-function regularization, it turns out that the only piece with a pole (the pole is at $s=1$) comes from the original 
\be
16\delta(j)\sum_{c=1}^\infty\frac{S(0,0;c)}{c}
\ee
This parallels our discussion in Appendix \ref{app:eisenstein}.

Now how about the extreme conical defect? (\ref{defect:label}) has density of state
\be
\rho_{ecd}(P,\bar{P})=\K^{(0,1)}_{P0}[\one] \bar{\K}^{(0,1)}_{\bar{P}0}[\one]
\ee
Now summing over modular transforms of the above gives
\be
\rho_{sum-ecd}(P,\bar{P})=\sum_{(c,d)=1}\K^{(c,d)}_{P0}[\one] \bar{\K}^{(c,d)}_{\bar{P}0}[\one]=\delta(P)\delta(\bar{P})+\sum_{\ell=-\infty}^\infty \sum_{c=1}^\infty \frac{8}{c}\delta(j-\ell)S(j,0;r)
\ee
Again the above sum is divergent and we can perform a zeta-function regularization to it. The only piece with a pole comes from the original term 
\be
8\delta(j)\sum_{c=1}^\infty \frac{S(0,0;c)}{c}
\ee
We can see it matches with the term of interest in $\rho_{MWK}$. More explicitly the zeta-function regularization gives
\begin{align}
&\sum_{\gamma\in SL(2,\Z)}\K^{(s,\gamma)}_{P0}[\one]\bar{\K}^{(s,\gamma)}_{\bar{P}0}[\one]=\delta(P)\delta(\bar{P})+4\sum_{\ell=-\infty}^\infty \sum_{c=1}^\infty \left(\frac{2\pi}{c}\right)^{2s}\frac{(P\bar{P})^{2s-1}}{\Gamma(s)^2}\delta(j-\ell)S(j,0;c)\\
&=\delta(P)\delta(\bar{P})+4(2\pi)^{2s}\frac{(P\bar{P})^{2s-1}}{\Gamma(s)^2}\delta(j)\frac{\zeta(2s-1)}{\zeta(2s)}+8\sum_{\ell=1}^\infty(2\pi)^{2s}\frac{(P\bar{P})^{2s-1}}{\Gamma(s)^2}\delta(j-\ell)\frac{\sigma_{1-2s}(\ell)}{\zeta(2s)}
\end{align}
Taking the limit $s\rightarrow\frac{1}{2}$, we have the regularized density of states
\be
\tilde{\rho}_{ed-sum}(P,\bar{P})=\delta(P)\delta(\bar{P})-\delta(P)\delta(\bar{P})+0=0
\ee
this agrees with expectation from $\tau$-basis analysis. 

%%%%
\section{Cusp one-loop determinants in 2d}
For JT gravity in two dimensions, the one loop determinant on the cusp is also proportional to a logarithmically divergent sum over images $\sum_{n\neq 0}^\infty \frac{1}{|n|}$. However, unlike what we found in three dimensions, in two dimensions the coefficient is zero, so the determinant is finite.

We write 2d hyperbolic space as
\be
\d s^2 = \frac{\d x^2 + \d y^2}{y^2}.
\ee
The double infinite cusp is the identification of this under $x \sim x + 1$. For JT gravity, the one loop determinant is \cite{Saad:2019lba}
\be
\frac{\det(-D_1+2)}{\det(-D_0+2)}
\ee
where
\be
D_n \equiv y^2(\partial_x^2 + \partial_y^2) - 2\i n y \partial_x.
\ee
The heat kernels for $(-D_0+m^2)$ and $(-D_1+m^2)$ are \cite{fay1977fourier}, see (5.17) of \cite{DHoker:1988pdl},
\begin{align}
K_0 &= \frac{\sqrt{2}e^{-(m^2+1/4)t}}{(4\pi t)^{3/2}}\int_r^\infty \d s\frac{s e^{-s^2/(4t)}}{\sqrt{\cosh(s)-\cosh(r)}}\\
K_1&=\frac{w'-\bar{w}}{w-\bar{w}'}\frac{\sqrt{2}e^{-(m^2+1/4)t}}{(4\pi t)^{3/2}}\int_r^\infty \d s\frac{s e^{-s^2/(4t)}}{\sqrt{\cosh(s)-\cosh(r)}}\left(2\frac{\cosh^2\frac{s}{2}}{\cosh^2\frac{r}{2}}-1\right).
\end{align}
Here $w = x+\i y$ and $r$ is the geodesic distance between $x,y$ and $x',y'$. For the sum over images we should set $x' = x+n$ and $y' = y$. Then $\cosh(r) = 1 + \frac{n^2}{2y^2}$. 

We want to compute
\be
\log \frac{\det(-D_1+2)}{\det(-D_0+2)}\supset \sum_{n\neq 0}\int_0^\infty \frac{\d t}{t}\int_0^\infty\frac{\d y}{y^2} (K_0-K_1).
\ee
But since the expression depends on $n$ and $y$ only through the combination $y/n$ the RHS is 
\be
\left(\sum_{n\neq 0}\frac{1}{|n|}\right)\int_0^\infty \frac{\d t}{t}\int_0^\infty\frac{\d y}{y^2} (K_0-K_1)|_{n = 1}.
\ee
This looks similar to the three dimensional case, in that we have a logarithmically divergent sum over $n$. But the difference is that the integrals over $t,y$ give zero (we checked this by doing the $t$ integral analytically and then the $s,y$ integrals numerically).

\bibliography{references}

\bibliographystyle{utphys}

\end{document}